\documentclass[10pt,a4paper,twocolumn,english,amssymb,nofootinbib,superscriptaddress,showpacs]{revtex4}
\usepackage{lmodern}

\usepackage[T1]{fontenc}
\usepackage[latin9]{inputenc}
\usepackage{color}
\usepackage{babel}
\usepackage{amsmath}
\usepackage{amssymb}
\usepackage{esint}
\usepackage[unicode=true,pdfusetitle,
 bookmarks=true,bookmarksnumbered=false,bookmarksopen=false,
 breaklinks=false,pdfborder={0 0 1},backref=false,colorlinks=false]
 {hyperref}

\makeatletter

\pdfpageheight\paperheight
\pdfpagewidth\paperwidth

\@ifundefined{textcolor}{}
{%
 \definecolor{BLACK}{gray}{0}
 \definecolor{WHITE}{gray}{1}
 \definecolor{RED}{rgb}{1,0,0}
 \definecolor{GREEN}{rgb}{0,1,0}
 \definecolor{BLUE}{rgb}{0,0,1}
 \definecolor{CYAN}{cmyk}{1,0,0,0}
 \definecolor{MAGENTA}{cmyk}{0,1,0,0}
 \definecolor{YELLOW}{cmyk}{0,0,1,0}
}

\usepackage{babel}
\@ifundefined{definecolor}{\usepackage{color}}{}
\@ifundefined{definecolor}{\usepackage{color}}{}
\usepackage{babel}
\@ifundefined{definecolor}{\usepackage{color}}{}
\usepackage{babel}

\pdfpageheight\paperheight
\pdfpagewidth\paperwidth

\@ifundefined{textcolor}{}{%
 \definecolor{BLACK}{gray}{0}
 \definecolor{WHITE}{gray}{1}
 \definecolor{RED}{rgb}{1,0,0}
 \definecolor{GREEN}{rgb}{0,1,0}
 \definecolor{BLUE}{rgb}{0,0,1}
 \definecolor{CYAN}{cmyk}{1,0,0,0}
 \definecolor{MAGENTA}{cmyk}{0,1,0,0}
 \definecolor{YELLOW}{cmyk}{0,0,1,0}
 }

\@ifundefined{definecolor}{\usepackage{color}}{}
\usepackage{latexsym}\usepackage{bm}
\usepackage{mathrsfs} % https://www.ctan.org/pkg/mathrsfs

\makeatother

\begin{document}

\title{Linearization Instability of Chiral Gravity }

\author{Emel Altas}

\email{altas@metu.edu.tr}

\selectlanguage{english}%

\affiliation{Department of Physics,\\
 Middle East Technical University, 06800 Ankara, Turkey}

\author{Bayram Tekin}

\email{btekin@metu.edu.tr}

\selectlanguage{english}%

\affiliation{Department of Physics,\\
 Middle East Technical University, 06800 Ankara, Turkey}
 
\begin{abstract}

Carrying out an analysis of the constraints and their linearizations on a spacelike hypersurface, we show that topologically massive gravity has a linearization instability at the chiral gravity limit about $AdS_3$. We also calculate the symplectic structure for all the known perturbative modes (including the log-mode) for the linearized field equations and find it to be degenerate (non-invertible) hence these modes do not approximate exact solutions and so do not belong  to the linearized phase space of the theory. Naive perturbation theory fails: the linearized field equations are necessary but not sufficient in finding viable linearized solutions. This has important consequences for both classical and possible quantum versions of the theory. 
\end{abstract}

\maketitle

\section{Introduction}
Quantum gravity is elusive not mainly because we lack computational tools, but because we do not know {\it what} to compute and so how to define the theory for a generic spacetime. One possible exception and a promising path is the case of asymptotically anti-de Sitter (AdS) spacetimes for which a dual  quantum conformal field theory that lives on the boundary of a bulk spacetime with gravity would amount to a definition of quantum gravity. But, even for this setting, we do not have a realistic four dimensional example. In three dimensions, the situation is slightly better: the cosmological Einstein's theory (with $\Lambda <0$) has a black hole solution \cite{BTZ} and possesses the right boundary symmetries (a double copy of the centrally extended Virasoro algebra \cite{BH}) for a unitary two dimensional conformal field theory. But as the theory has no local dynamics (namely gravitons), it is not clear exactly how much one can learn from this model as far as quantum gravity is concerned. Having said that, even for this ostensibly simple model, we still do not yet have a quantum gravity theory. Recasting Einstein's gravity in terms of a solvable Chern-Simons gauge theory is a possible avenue \cite{Witten88}, but this only works for non-invertible dreibein  which cannot be coupled to generic matter. 

A more realistic gravity in three dimensions is the topologically massive gravity (TMG) \cite{djt} which has black hole solutions as well as a dynamical massive graviton. But the apparent problem with TMG is that the bulk graviton and the black hole cannot be made to have positive energy generally.  This obstruction to a viable classical and perhaps quantum theory was observed to disappear in an important work \cite{Strom1}, where it was realized that at a "chiral point"  defined by a tuned topological mass in terms of the AdS radius, one of the Virasoro algebras has a vanishing central charge (and so admits a trivial unitary representation) and the other has a positive nonzero central charge with unitary nontrivial representations,  the theory has a positive energy black hole and zero energy bulk gravitons. This tuned version of TMG, called "chiral gravity", seems to be a viable candidate for a well-behaved classical and quantum gravity. 

One of the main objections raised against the chiral gravity is that it possesses a negative energy perturbative log-mode about the AdS vacuum which ruins the unitarity of the putative boundary CFT \cite{Grumiller}. Of course if this is the case, chiral gravity is not even viable at the classical level, since it does not have a vacuum.  It was argued in \cite{Strom2,Carlip} that chiral gravity could survive if the theory is linearization unstable about its AdS solution. This means that there would be perturbative modes which cannot be obtained from any exact solution of the theory.  In fact, these arguments were supported with the computations given in \cite{emel} where it was shown that the Taub charges which are functionals quadratic in the perturbative modes that must vanish identically due to background diffeomorphism invariance, do not vanish for the log-mode that ruins the chiral gravity.  This means that the log-mode found from the linearized field equations is an artifact of the linearized equations and does not satisfy the global constraints coming from the Bianchi identities. 

In this work, we give a direct proof of the linearization instability of chiral gravity in AdS using the constraint analysis of the full TMG equations defined on a spacelike hypersurface. The crux of the argument that we shall lay out below is the following: the linearized constraint equations of TMG show that there are inconsistencies exactly at the chiral point. Namely perturbed matter fields do not determine the perturbations of the metric components on the  spacelike hypersurface and there are unphysical constraints on matter perturbations besides the usual covariant conservation. 

To support our local analysis on the hypersurface, we compute  the symplectic structure (that carries all the information about the phase space of the theory) for all perturbative solutions of the linearized field equations and find that the symplectic 2-form is degenerate and so non-invertible hence these modes do not approximate ({\it i.e.}\thinspace they are not tangent to) actual nonlinear solutions.
The symplectic 2-form evaluated for the log-mode is time-dependent (hence not coordinate-invariant) and vanishes at the initial value surface and grows unbounded in the future. 

To carry out the constraint analysis and their linearizations (which will yield possible nearby solutions to exact solution), we shall use the field equations instead of the TMG action as  the latter is not diffeomorphism invariant which complicates the discussion via the introduction of tensor densities (momenta) instead of tensors. We shall also work in the metric formulation instead of the first order one as there can be significant differences between the two formulations. Before we indulge into the analysis, let us note that the linearization instability that arises in the perturbative treatment of nonlinear theories and can be confused with dynamical or structural instability, as both are determined with the same linearization techniques.The difference is important: the latter refers to a real instability of a system such as the instability of the vacuum in a theory with ghosts such as the $R+\beta R_{\mu\nu}^2$ theory with $\beta\neq0$, this is simply not physically acceptable. On the other hand linearization instability refers to the failure of perturbation theory for a given background solution and one should resort to another method to proceed. From the point of view of the full solution space of the theory, this means that this (possibly infinite dimensional) space is not a smooth manifold but it has conical singularities around certain solutions. Let us expound on this a little bit.

\section{linearization instability in brief }

A nonlinear equation $F(x)=0$  is said to be linearization stable at a solution $x_0$ if every solution $\delta x$ to the linearized equation $F^\prime(x_{0})\cdot\delta x=0 $ is tangent to a curve of solutions to the original nonlinear equation. In some nonlinear theories, not all solutions to the linearized field equations represent linearized versions of exact (nonlinear) solutions. As a common algebraic example, let us consider the function $F(x,y)= x( x^2 + y^2)=0$, where $x,y$ are real, exact solution space is one dimensional given as $(0, y)$, and the linearized solution space is also one dimensional $(0,  \delta y ) $ as long as $y\ne0$. But at exactly the solution $(0, 0)$, the linearized solution space is two dimensional $(\delta x,  \delta y )$ and  so there are clearly linerized solutions with $\delta x \ne 0$, which do not come from the linearization of any exact solution. The existence of such spurious solutions depends on the particular theory at hand and the background solution (with its symmetries and topology) about which linearization is carried out.  If such so called "nonintegrable" solutions exist,  perturbation theory in some directions of solution space fails and we say that the theory is not linearization stable at a nonlinear exact solution. 

What we have just described is not an exotic phenomenon: a {\it priori} no nonlinear theory is immune to linearization instability: one must study the problem case by case. For example, pure general relativity is linearization stable in Minkowski spacetime (with a non-compact Cauchy surface) \cite{Choquet_Deser}, hence perturbation theory makes sense, but it is not linearization stable on a background with compact Cauchy surfaces that possesses at least one Killing symmetry \cite{Moncrief} which is the case when the Cauchy surface is a flat 3-torus \cite{Deser_Brill}: on $T^3\times R$, at second order of the perturbation theory, one must go back and readjust the first order perturbative solution.

As gravity is our main interest here, let us consider some nonlinear gravity field equations  in a coordinate chart as $\mathscr{E}_{\mu\nu}=0$, which admits $\bar{g}_{\mu\nu}$ as an exact solution, if {\it every} solution ${h}_{\mu\nu}$ of the linearized field equations $\mathscr{E}^{(1)}(\bar{g})\cdot h=0$ is tangent to an exact solution ${g}_{\mu\nu}(\lambda)$ such that ${g}_{\mu\nu}(0)=\bar{g}_{\mu\nu}$ and $\frac{dg_{\mu\nu}}{d\lambda}|_{\lambda=0}=h_{\mu\nu}$ then, according to our definition above, the theory is linearization stable. Otherwise it is linearization unstable. In general, we do not have a theorem stating the {\it necessary and sufficient} conditions  for the linearization stability of a generic gravity theory about a given exact solution. For a detailed discussion on generic gravity models, see our recent work \cite{emel}. But, as discussed in section II of that work, defining the second order perturbation as $\frac{d^2 g_{\mu\nu}}{d\lambda^2}|_{\lambda=0}=k_{\mu\nu}$, if the following second order equation
\begin{equation} 
(\mathscr{E})^{(2)}(\bar{g})\cdot [h,h]+(\mathscr{E})^{(1)}(\bar{g})\cdot k=0,
\label{lin1}
\end{equation}
has a solution for $k_{\mu \nu}$ without a constraint on the linear solution $h_{\mu\nu}$, then the theory 
is linearization stable. Of course, at this stage it is not clear that there will arise no further constraints on the linear theory beyond the second order perturbation theory. In fact, besides Einstein's theory, this problem has not been worked out. But in Einstein's gravity, as the constraint equations are related to the zeros of the moment map, one knows that there will be no further constraint for the linear theory coming from higher order perturbation theory beyond the second order  \cite{Marsden_lectures}. In Einstein's gravity for compact Cauchy surfaces without a boundary, the necessary and sufficient conditions are known for linearization stability \cite{Moncrief,M1,M2,M3}.  

In practice, it is very hard to show that (\ref{lin1}) is satisfied for {\it all } linearized solutions, therefore,  one resorts to a weaker condition by contracting that equation with a Killing vector field and integrates over a hypersurface to obtain  $ Q_{Taub}\left[\bar{\text{\ensuremath{\xi}}}\right] +Q_{ADT}\left[\bar{\text{\ensuremath{\xi}}}\right] =0$
where the Taub charge \cite{Taub} is defined as\footnote{As it appears in the second order perturbation theory, the Taub charge is not a widely known quantity in physics, for a more detailed account of it, we invite the reader to study the relevant section of \cite{emel}} 
\begin{equation}
Q_{Taub}\left[\bar{\text{\ensuremath{\xi}}}\right]:=\intop_{\Sigma} d^{3}\Sigma\thinspace\sqrt{\gamma}\thinspace\hat{n}^{\nu}\thinspace\bar{\text{\ensuremath{\xi}}}^{\mu}\thinspace(\text{\ensuremath{\mathscr{E}}}_{\mu\nu})^{(2)}\cdot [h,h],
\label{ttt}
\end{equation}
and the ADT charge \cite{Abbott_Deser,adt} is defined as
\begin{equation}
Q_{ADT}\left[\bar{\text{\ensuremath{\xi}}}\right] :=\intop_{\Sigma} d^{3}\Sigma\thinspace\sqrt{\gamma}\thinspace\hat{n}^{\nu}\thinspace\bar{\text{\ensuremath{\xi}}}^{\mu}\left(\text{\ensuremath{\mathscr{E}}}_{\mu\nu}\right)^{(1)}\cdot k.
\label{ADT}
\end{equation}
The latter can be expressed as a boundary integral. For the case of compact Cauchy surfaces without a boundary, $Q_{ADT} =0$, and hence one must have $Q_{Taub}=0$ which leads to the aforementioned quadratic integral constraint on the linearized perturbation $h_{\mu\nu}$ as the integral in (\ref{ttt}) should be zero. This is the case for Einstein's gravity, for example, on a flat 3-torus: $Q_{Taub}$ does not vanish automatically and so the first order perturbative result $h$ is constrained. On the other hand, for extended gravity theories (such as the theory we discuss here), $Q_{ADT}$ vanishes for a different reason, even for non-compact surfaces, as in the case of AdS. The reason is that for some tuned values of the parameters in the theory, the contribution to the conserved charges from various tensors cancel each other exact, yielding nonvacuum solutions that carry the (vanishing) charges of the vacuum. This is the source of instability.

\section{ADM decomposition of TMG}

Before restricting to the chiral gravity limit, we first study the full TMG field equations coupled with matter fields as an initial value problem, hence we take 
\begin{equation}
\mathscr{E}_{\mu\nu}=G_{\mu\nu}+\Lambda g_{\mu\nu}+\frac{1}{\mu}C_{\mu\nu}=\kappa\tau_{\mu\nu}.
\end{equation}
 The ADM \cite{ADM} decomposition of the metric reads
\begin{equation}
ds^{2}=-(n^{2}-n_{i}n^{i})dt^{2}+2n_{i}dtdx^{i}+\gamma_{ij}dx^{i}dx^{j},
\end{equation}
where ($n$, $n_{i}$) are lapse and shift  functions and $\gamma_{ij}$ is the $2D$
spatial metric. From now on, the Greek indices will run over the full spacetime, while the Latin indices will run over the  hypersurface  $\varSigma$, as $i,j...=1,2$. The spatial indices will be raised and lowered by the $2D$ metric. The extrinsic curvature ($k_{i j}$) of the surface is given as 
\begin{equation}
2 n k_{ij}=\dot{\mathbf{\gamma}}_{ij}-2 D_{(i}n_{j)},
\end{equation}
where $D$ is the covariant derivative compatible with $\gamma_{ij}$ and $\dot{\mathbf{\gamma}}_{ij} :=\partial_{0}\gamma_{ij}$ and the round brackets denote symmetrization with a factor of 1/2. With the convention 
 $R_{\rho\sigma}=\partial_{\mu}\Gamma_{\rho\sigma}^{\mu}-\partial_{\rho}\Gamma_{\mu\sigma}^{\mu}+\Gamma_{\mu\nu}^{\mu}\Gamma_{\rho\sigma}^{\nu}-\Gamma_{\sigma\nu}^{\mu}\Gamma_{\mu\rho}^{\nu}$, one finds  the hypersurface components of the three dimensional Ricci tensor as 
\begin{eqnarray}
&&R_{ij}={^{(2)}R}_{ij}+k k_{ij}-2k_{ik}k_{j}^{k}  \\ 
&&+\frac{1}{n}(\dot{k}_{ij}
-n^{k}D_{k}k_{ij}  -D_{i}\partial_{j}n-2 k_{k(i}D_{j)}n^{k}), \nonumber
\end{eqnarray}
where ${^{(2)}R}_{ij}$ is the Ricci tensor of the hypersurface and $k\equiv \gamma^{i j} k_{ij}$. Similarly one find the twice projection to the normal of the surface as 
\begin{align}
R_{00}=&\frac{n^{i}n^{j}}{n}(\dot{k}_{ij}-n^{k}D_{k}k_{ij}-D_{i}\partial_{j}n-2k_{kj}D_{i}n^{k}) \nonumber\\&-n^{2}k_{ij}^{2}
+n^{i}n^{j}(^{(2)}R_{ij}+k k_{ij}-2k_{ik}k_{j}^{k})\\&+n(D_{k}\partial^{k}n-\dot{k}-n^{k}D_{k}k+2n^{k}D_{m}k_{k}^{m}). \nonumber
\end{align}
On the other hand, projecting once to the surface and once normal to the surface yields 
\begin{align}
R_{0i}&=\frac{n^{j}}{n}(\dot{k}_{ij}-n^{k}D_{k}k_{ij}-D_{i}\partial_{j}n-2k_{k(i}D_{j)}n^{k})\\&+n^{j} ({^{(2)}R}_{ij}+kk_{ij}-2k_{ik}k_{j}^{k})+n(D_{i}k+D_{m}k_{i}^{m}). \nonumber
\end{align}
We also need the 3D scalar curvature in terms of the hypersurface quantities which can be found as 
\begin{equation}
R={}^{(2)}R+k^{2}-k_{ij}^{2}+\frac{2}{n}(\dot{k}+n k_{ij}^{2}-D_{i}D^{i}n-n^{i}D_{i}k).
{\label{3DR}}
\end{equation}
Given the Schouten tensor $S_{\mu\nu}:=R_{\mu\nu}-\frac{1}{4}Rg_{\mu\nu}$, the Cotton tensor is defined as
\begin{equation}
C_{\mu\nu} :=\frac{1}{2}\epsilon{}^{\rho\alpha\beta}(g_{\mu\rho}\nabla_{\alpha}S_{\beta\nu}+g_{\nu\rho}\nabla_{\alpha}S_{\beta\mu}),
\end{equation}
where  $\epsilon{}^{\rho\alpha\beta}$ is the totally antisymmetric tensor which splits as $\epsilon{}^{0mn}=\frac{1}{n}\epsilon^{mn}=\frac{1}{n}\gamma^{-\frac{1}{2}}\varepsilon^{mn}$ where $\varepsilon^{mn}$ is the antisymmetric symbol. Just as we have done the ADM decomposition of the Ricci tensor, a rather lengthy computation yields the following expressions, for the projections of the Cotton tensor
\begin{align}
2 n C_{ij}=&\epsilon{}^{mn}n_{i}( D_{m}S_{nj}-k_{mj}(D_{r}k_{n}^{r}-\partial_{n}k)) \nonumber\\
&+\epsilon{}^m\,_i\bigg \{\dot{S}_{mj}-n k_{j}^{k}S_{mk}-S_{mk}D_{j}n^{k} \nonumber \\
&-(\partial_{j}n+n^{r}k_{rj})(D_{s}k_{m}^{s}-\partial_{m}k) \nonumber \\
&-D_{m}(n^{r}S_{rj}+n(D_{r}k_{j}^{r}-D_{j}k) ) \nonumber \\
&+k_{mj}(D_{k}\partial^{k}n-\dot{k}+n^{k}D_{s}k_{k}^{s}+n(\frac{R}{4}-k_{rs}^{2})) \bigg \} \nonumber \\
&+i\leftrightarrow j,
\end{align}
and 
\begin{equation}
C_{i0}=n^{j}C_{ij}-\frac{\epsilon{}^{mn}}{2}(n A_{mni}-n_{i}B_{mn}-\gamma_{in}( C_{m}+n E_{m}))
\end{equation}
and
\begin{align}
&C_{00}=n^{i}n^{j}C_{ij} \\
&-\epsilon{}^{mn}(nn^{i}A_{mni}-(n_{i}n^{i}-n^{2})B_{mn}-n_{n}(C_{m}+n E_{m})),
\nonumber
\end{align}
where we have defined the following tensors
\begin{align*}
&A_{mni}\equiv D_{m}S_{ni}-k_{mi}\left(D_{r}k_{n}^{r}-\partial_{n}k\right), \\
&B_{mn}\equiv D_{m}D_{r}k_{n}^{r}-k_{m}^{k}S_{kn}, \\
&E_{m}\equiv 2k_{rs}D_{m}k^{rs}-\frac{1}{4}\partial_{m}R+k_{m}^{k}\left(D_{r}k_{k}^{r}-\partial_{k}k\right), \\
&C_{m}\equiv \partial_{0}D_{r}k_{m}^{r}-S_{m}^{k}\left(\partial_{k}n+n^{r}k_{rk}\right)-D_{m}D_{k}\partial^{k}n\\
&\,\,\,\,\,\,\,-D_{m}\left(n^{k}D_{s}k_{k}^{s}\right)+k_{m}^{k}S_{kr}n^{r}+\partial_{m}n(k_{rs}^{2}-\frac{R}{4}).
\end{align*}
Using  the above decomposition, we can recast the ADM form of the full TMG equations as
\begin{equation}
\mathscr{E}_{ij}=S_{ij}-\frac{1}{4}\gamma_{ij}R+\Lambda\gamma_{ij}+\frac{1}{\mu}C_{ij}=\kappa\tau_{ij}
\end{equation}
and
\begin{align}
\mathscr{E}_{0i}=&\kappa\tau_{0i}=n^{j}\mathscr{E}_{ij}+n(D_{r}k_{i}^{r}-\partial_{i}k)\\-&\frac{1}{2\mu}\epsilon{}^{mn}(n A_{mni}-n_{i}B_{mn}-\gamma_{in}(C_{m}+n E_{m})) \nonumber
\end{align}
and
\begin{align}
\mathscr{E}_{00}&=\kappa\tau_{00}=2n^{i}\mathscr{E}_{0i}-n^{i}n^{j}\mathscr{E}_{ij}-\Lambda n^{2}-\frac{1}{\mu}\epsilon{}^{mn}n^{2}B_{mn}\nonumber \\&+n(D_{k}\partial^{k}n-\dot{k}+n^{k}D_{k}k+n(\frac{R}{2}-k_{rs}^{2})).
\end{align}
From $\mathscr{E}_{0i}$, we get the momentum constraint as
\begin{align}
\Phi_{i}&=\kappa(\tau_{0i}-n^{j}\tau_{ij})=n(D_{r}k_{i}^{r}-\partial_{i}k)\\+&\frac{1}{2\mu}\epsilon{}^{mn}(n_{i}B_{mn}-nA_{mni}+\gamma_{in}C_{m}+n\gamma_{in}E_{m}) \nonumber
\end{align}
and from $\mathscr{E}_{00}$ we get the Hamiltonian constraint as 
\begin{align}
\Phi=&\frac{\kappa}{n^{2}}(\tau_{00}-2n^{i}\tau_{0i}+n^{i}n^{j}\tau_{ij}) \nonumber \\
+&\frac{1}{2}(^{(2)}R+k^{2}-k_{ij}^{2}-2\Lambda) \nonumber \\
 -&\frac{1}{\mu}\epsilon{}^{mn}\left(D_{m}D_{r}k_{n}^{r}-k_{m}^{k}S_{kn}\right),
\end{align}
where in the last equation we made use of the explicit form of $R$ given in (\ref{3DR}) which for TMG is  $R=6\Lambda-2 \kappa\tau $. From now on, for our purposes, it will suffice to work in the Gaussian normal coordinates with $n=1$ and $n_{i}=0$ for which $k_{ij}=\frac{1}{2}\dot{\gamma}_{ij}$ and the constraints reduce to
\begin{align}
&\frac{\epsilon^{mn}}{4\mu}(\dot{\gamma}_{i m}\gamma^{i k}(^{(2)}R_{kn}-\dot{\gamma}_{kp}\dot{\gamma}_{sn}\gamma^{ps}-\ddot{\gamma}_{kn})-2 D_{m}D^{k}\dot{\gamma}_{kn}) \nonumber \\
&-\frac{1}{8}\dot{\gamma}_{ij}\left(\dot{\gamma}_{ab}\gamma^{ab}\gamma^{ij}+\dot{\gamma}^{ij}\right)=\kappa\tau_{00}+\Lambda-\frac{^{(2)}R}{2}
\end{align}
and
\begin{align}
&\frac{\epsilon^{m}\thinspace_{i}}{8\mu}\left( \dot{\gamma}^{kp}(2 D_{k}\dot{\gamma}_{pm}-D_{m}\dot{\gamma}_{kp})+2D^{k}\ddot{\gamma}_{km}-\dot{\gamma}_{mk}\gamma^{kl}D^{p}\dot{\gamma}_{pl}\right) \nonumber\\
&-\frac{\epsilon^{mn}}{8\mu}\bigg(\dot{\gamma}_{ab}\gamma^{ab}D_{m}\dot{\gamma}_{in}-2\gamma^{ks}D_{m}(\dot{\gamma}_{kn}\dot{\gamma}_{si}) \nonumber  \\
&+2D_{m}\ddot{\gamma}_{in}-\dot{\gamma}_{mi}D^{k}\dot{\gamma}_{kn}\bigg) \\
&+\frac{1}{2}\left(D^{k}\dot{\gamma}_{ki}-\gamma^{ab}D_{i}\dot{\gamma}_{ab}\right)=\kappa\tau_{0i}+\frac{1}{2\mu}\epsilon^{mn}D_{m}{}^{(2)}R_{ni}. \nonumber
\end{align}
Furthermore, taking a conformally flat $2D$ metric on $\Sigma$, we have $\gamma_{ij}=e^{\varphi}\delta_{ij}$, where $\varphi=\varphi(t,x_i)$, $k_{ij}=\frac{1}{2}\dot{\varphi}\gamma_{ij}$ and the $2D$ Ricci tensor becomes
\begin{equation}
^{(2)}R_{ij}=-\frac{1}{4}\gamma_{ij}e^{-\varphi}\left(2D_{k}\partial_{k}\varphi+\partial_{k}\varphi\partial_{k}\varphi\right),
\end{equation}
whereas the $3D$ Ricci tensor reads
\begin{equation}
R_{ij}=\frac{1}{2}\gamma_{ij}(-D^{k}\partial_{k}\varphi+\dot{\varphi}^{2}+\ddot{\varphi}-\frac{1}{2}\partial^{k}\varphi\partial_{k}\varphi)
\end{equation}
and the $3D$ scalar curvature is 
\begin{equation}
R=-D^{k}\partial_{k}\varphi+\frac{3}{2}\dot{\varphi}^{2}+2\ddot{\varphi}-\frac{1}{2}\partial^{k}\varphi\partial_{k}\varphi.
\end{equation}
With all these results in hand, one can obtain from the constraint equations the following relation
\begin{equation}
\partial_{i}\dot{\varphi}=-J_{i}+\frac{1}{2\mu}\epsilon^{m}\thinspace_{i}\dot{\varphi}\partial_{m}\dot{\varphi},
\label{kolay}
\end{equation}
where we have introduced the "source current"  which, on the hypersurface, reads
\begin{equation}
J_{i} := 2\kappa\tau_{0i}+\frac{\kappa}{\mu}\epsilon^{m}\thinspace_{i}\partial_m\tau_{00}.
\end{equation}
Contracting (\ref{kolay}) with the epsilon-tensor, one arrives at
 \begin{equation}
\frac{2\mu}{\dot{\varphi}}\epsilon^{mi}\partial_{m}\dot{\varphi}\left(1+\frac{\dot{\varphi}^{2}}{4\mu^{2}}\right)=-\frac{2\mu}{\dot{\varphi}}\epsilon^{mi}J_{m}+J^{i}.
{\label{main_equation}}
\end{equation}
In the case of vacuum, $\tau_{\mu\nu}=0$, and so $J_{i}=0$,  the unique solution to (\ref{main_equation}) is of the form $\varphi_{0}= c t$, where $c $ is a constant which can be found from the trace equation that reads  $R=6\Lambda$. So $\ensuremath{c}=2\sqrt{\Lambda} \equiv \frac{2}{\ell}$, which is the de Sitter (dS) solution and $\ell >0$ is its radius. Turning on a compactly supported matter perturbation with $\delta\tau_{\mu\nu}\ne0$, one has $\delta J_i\ne0$ and perturbing the constraint equations about $\varphi_{0}$ as $\varphi=\varphi_{0}+\delta\varphi$, we find a linearized constraint equation 
\begin{align}\label{pert}
&\mu(1+\frac{1}{\mu^{2} \ell^2})\epsilon^{m}\thinspace_{i}\partial_{m}\delta\dot{\varphi}  \\=
&(\partial_{i}+\frac{1}{\mu \ell}\epsilon^{m}\thinspace_{i}\partial_{m})\kappa\delta\tau_{00}+2\mu(\epsilon_{i}\thinspace^{m}+\frac{1}{\mu \ell}\delta^{m}\thinspace_{i})\kappa\delta\tau_{0m}\nonumber,
\end{align}
from which, for the dS case, one can solve the perturbation ($\delta \varphi$) and hence the perturbed metric  by integration
in terms of the perturbed matter fields  on the  hypersurface. Hence dS is linearization stable in TMG for any finite value of $\mu \ell$. The other linearized constraints are compatible with this solution. Our computation has been analytic in $\ell$, hence, we can do the following "Wick" rotation to study the  AdS case: $x^i\rightarrow ix^i$, $t\rightarrow it$, $ \ell \rightarrow i \ell$ yielding  $\Lambda=-\frac{1}{\ell^{2}}$ with the Gaussian normal form of the (signature changed) metric 
$ds^{2}=dt^{2}-e^{-{2 t/ \ell}}\left(dx^{2}+dx^{2}\right).$ Then for AdS,  (\ref{pert}) becomes 
\begin{align}\label{pert2}
&\mu (1-\frac{1}{\mu^{2} \ell^2})\epsilon^{m}\thinspace_{i}\partial_{m}\delta\dot{\varphi}  \\=
&-(\partial_{i}-\frac{1}{\mu \ell}\epsilon^{m}\thinspace_{i}\partial_{m})\kappa\delta\tau_{00}-2\mu (\epsilon_{i}\thinspace^{m}+\frac{1}{\mu \ell}\delta^{m}\thinspace_{i})\kappa\delta\tau_{0m}\nonumber
\end{align}
and once again the perturbation theory is valid for {\it generic } values of $\mu \ell$ in AdS as in the case of dS. But at the chiral point, $\mu \ell =1$,  the left-hand side vanishes identically and there is an unphysical constraint on the matter perturbations $\delta\tau_{0m}$ and $\delta\tau_{00}$ in addition to their background covariant conservation. Moreover, the metric perturbation is not determined by the matter perturbation. What this says is that in the chiral gravity limit of TMG, for AdS, the exact AdS solution is linearization unstable. The above computation has been a local one, and does not depend on the fact that AdS does not have a Cauchy surface on which one can define the initial value problem. AdS requires initial and boundary values together, but what we have computed is a necessary condition for such a formulation (not a sufficient one) and AdS in chiral gravity does not satisfy the necessary conditions for the initial-boundary value problem.

\section{Symplectic structure of TMG}
Let us give another argument for the linearization instability of AdS making use of the symplectic structure of TMG which was found in \cite{caner} following \cite{w} as 
$\omega := \int_\Sigma d \Sigma_\alpha  \sqrt{|g|} {\cal{J}}^\alpha$,
where $\Sigma$ is the hypersurface. $\omega$ is a closed ($\delta w=0$)  non-degenerate (except for gauge directions) 2-form for full TMG including chiral gravity.  Here the on-shell covariantly conserved symplectic current reads
\begin{align}
 {\cal{J}}^\alpha&=\delta  \Gamma^\alpha_{\ \mu\nu} \wedge  ( \delta  g^{\mu \nu} + \frac{1}{2}  g^{\mu \nu} \delta \ln g ) \nonumber \\
&-  \delta  \Gamma^\nu_{\ \mu\nu} \wedge  ( \delta  g^{\alpha \mu} + \frac{1}{2}  g^{\alpha \mu} \delta \ln g ) \nonumber \\
		&+   \frac{1}{\mu}\epsilon^{\alpha \nu \sigma} ( \delta  {S}^\rho_{\ \sigma} \wedge \delta g_{\nu \rho}
					+ \frac{1}{2} \delta \Gamma^\rho_{\ \nu \beta} \wedge  \delta \Gamma^\beta_{\ \sigma \rho}  ).
\label{symplectic_two}
\end{align}
What is important to understand is that  $\omega$ is a gauge invariant object on the solution space, say ${\mathcal {Z}}$, and also on the (more relevant) quotient ${\mathcal {Z}}/Diff$ which is the phase space and $Diff$ is the group of diffeomorphisms. Therefore, even without knowing the full space of solutions, by studying the symplectic structure, one gains a lot of information for both classical and quantum versions of the theory. Perturbative solutions live in the tangent space of the phase space and hence they are crucial in the discussion. We refer the reader to \cite{caner} for a full discussion of this.

Let us show that for the linearized solutions of chiral gravity given in \cite{Strom1}   the symplectic 2-form is degenerate and hence not invertible. In the global coordinates, the background metric reads
\begin{equation}
 ds^2 = \ell^2\big(-\cosh^2{\rho}\, d\tau^2 +\sinh^2{\rho}\,d\phi^2+d\rho^2\big),
\end{equation}
defining  $u=\tau+\phi$, $v=\tau-\phi$, making use of the $SL(2, R)\times SL(2, R)$, \cite{Strom1} found all  the primary states (but one) and their descendants. The primary solutions are
\begin{equation}
h_{\mu \nu} = \Re \left\{e^{ -i  \Delta \tau -i S \phi } F_{\mu \nu}(\rho)\right\},
\end{equation}
where the real part is taken and the background tensor reads
\begin{eqnarray}
F_{\mu\nu}(\rho)=f(\rho)\left(\begin{array}{ccc}
                                         1 & {S\over2}& {2i\over\sinh 2\rho} \\
                                          {S\over2} & 1 & {i S\over\sinh 2\rho} \\
                                         {2i\over\sinh 2 \rho} &{i S\over \sinh 2 \rho}   & - {4\over\sinh^2 2\rho} \\
                                       \end{array}\right)
\end{eqnarray}
and $f(\rho)=(\cosh{\rho})^{-\Delta }\sinh^2{\rho}$,
where $\Delta \equiv  h+\bar{h}$ and $S \equiv  h-\bar{h}$.
 Components of the symplectic current for these modes (for generic $\mu \ell$) can be found as
\begin{align}
 &{\cal{J}}^\tau =\frac{( 4-S^2)(S + 2 \mu \ell)\Delta}{8 \mu \ell^7 (\cosh\rho)^{2(1+\Delta)}}\sin \left( 2 \Delta \tau +2 S \phi \right), \nonumber \\
& {\cal{J}}^\phi= -\frac{2 \coth^2 \rho}{S+ 2 \mu \ell} {\cal{J}}^\tau, \\
&{\cal{J}}^\rho =-\frac{(S \Delta +4\mu \ell)\coth \rho +(\Delta -2)\mu \ell \sinh 2 \rho}
{\Delta ( S+ 2 \mu \ell)}{\cal{J}}^\tau, \nonumber
\end{align}
which yield a vanishing  $\omega$ at the chiral limit since for left, right and massive modes we have $S^2 = 4$ and the relevant symplectic current ${\cal{J}}^\tau$ vanishes identically, hence the solution is not viable. Moreover, one can show that its Taub charge diverges, while its ADT charge is for the background Killing vector $(-1,0,0)$ is
\begin{eqnarray}
Q_{ADT}=&&-\lim_{r \rightarrow \infty}\frac{ \sin (\pi S) \cos ( 2  \pi S + \Delta t)}{ 4 \pi S 2^{2-\Delta}\ell}\Delta ( 2 \Delta + S-2) \nonumber \\
&&\times  e^{ r(2-\Delta)},  
\end{eqnarray}
which vanishes for the massive mode $\Delta = S=2$. 
In addition to the above solutions, there is an additional  the log-mode given in \cite{Grumiller} which reads
\begin{align}
 &h_{ \mu\nu} = f_1(\tau,\rho)\left( \begin{array}{ccc}
0 & 0 & 1 \\
0 & 0 & 1 \\ 
1 & 1 & 0
\end{array} \right)_{\mu\nu} \nonumber \\
&+f_2(\tau, \rho)\left( \begin{array}{ccl}
1 & 1 & \qquad \qquad 0 \\
1 & 1 & \qquad \qquad 0 \\
0 & 0 & - {4\over\sinh^2 2\rho}
\end{array} \right)_{\mu\nu},
\label{grumillermetric}
\end{align}where the two functions are given as 
\begin{eqnarray}
&&f_1(\tau, \rho) =  \frac{\sinh{\rho}}{\cosh^3{\rho}}\,(\tau \cos{2u}-\sin{2u}\,\ln{\cosh{\rho}}) ,\nonumber \\
&&f_2(\tau, \rho) =-\tanh^2\!\!{\rho}\,(\tau \sin{2u}+\cos{2u}\,\ln{\cosh{\rho}}). \nonumber 
\end{eqnarray}
The components of the symplectic current for this mode  read
\begin{align}
 &{\cal{J}}^\tau = \frac{1}{\mu \ell^7}\tau ((1-\mu \ell ) \cosh 2 \rho+1)\text{sech}^{10}\rho , \nonumber \\
& {\cal{J}}^\phi= - \frac{2}{\mu \ell^7}\tau (1- \mu \ell) \text{sech}^8 \rho, \\
&{\cal{J}}^\rho=\frac{1}{ \ell^6}\tanh \rho\, \text{sech}^8\rho(4
   (\log ^2\cosh \rho+\tau ^2)+\log \text{sech}\,\rho) \nonumber ,
\end{align}
which yield a linearly growing $\omega$ in $\tau$ and vanishes on the initial value surface.
 What all these say is that first order perturbation theory simply fails in chiral gravity limit of TMG. If the theory makes any sense at the classical and/or quantum level one must resort to a new method to carry out computations.  This significantly affects its interpretation in the context of AdS/CFT as the perturbed metric couples to the energy-momentum tensor of the  boundary CFT. This of course does not say anything about the solutions of the theory which are not globally AdS and one might simply have to define the theory in a different background.

\section{Conclusions}

The problem studied here is a frequently recurring one \cite{kastor}, for example it also appears in critical gravity \cite{pope, sisman22}. Linearized solutions by definition satisfy the linearized equations but this is not sufficient; they should also satisfy a quadratic constraint to actually represent linearized versions of exact solutions. This deep result comes from the Bianchi identities and their linearizations and it is connected to the conserved quantities. With the observation of gravity waves, research in general relativity and its modifications, extensions has entered an exciting era in which many theories might be possibly tested. One major tool of computation in nonlinear theories, such as gravity, is perturbation theory from which one obtains a lot of information and the gravitational wave physics is no exception as one uses the tools of perturbation theory to obtain the wave profile far away from the sources. Therefore, the issue of linearization instability arises in any use of perturbation theory as the examples provided here and before \cite{emel} show even for the ostensibly safe case of spacetimes with noncompact Cauchy surfaces.

\end{document}